\documentclass[prl,twocolumn,showpacs,superscriptaddress]{revtex4}

\usepackage{amsmath,amssymb,graphicx}

\begin{document}

\title{Single File Diffusion enhancement in a fluctuating modulated 1D channel}
\date{\today}
\author{Gwennou Coupier}\author{Michel Saint Jean}\email{michel.saintjean@paris7.jussieu.fr}\author{Claudine Guthmann}
 \affiliation{Laboratoire
Mati\`ere et Syst\`emes Complexes, UMR 7057 CNRS \& Universit\'e
Paris~7 - 140 rue de Lourmel, F-75015 Paris, France}

\begin{abstract}
We show that the diffusion of a single file of particles moving in a
fluctuating modulated 1D channel is enhanced with respect to the one
in a bald pipe. This effect, induced by the fluctuations of the
modulation, is favored by the incommensurability between the channel
potential modulation and the moving file periodicity. This
phenomenon could be of importance in order to optimize the critical
current in superconductors, in particular in the case where mobile
vortices move in 1D channels designed by adapted patterns of pinning
sites.

\end{abstract}
 \maketitle

\begin{figure*}
\resizebox{2\columnwidth}{!}{\includegraphics{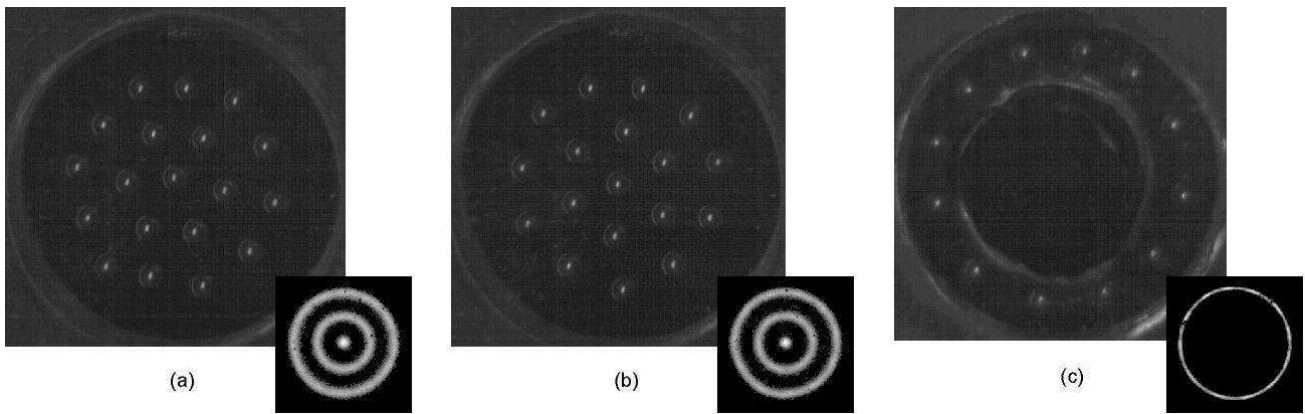}}
 \caption{\label{fig:fig1} Experimental configurations. Wigner
islands : (a) $N=19$, (b) $N=18$ ; circular bald pipe (c) $N=12$. In
inset, the corresponding balls trajectories at the same effective
temperature.}
\end{figure*}

Recent progress in lithography techniques have given rise to many
developments to obtain superconducting mesoscopic devices with large
critical currents. These currents being dependent upon the vortices
pinning, many studies have been focused in the last past years to
understand the properties of a mobile file of vortex moving in 1D
channels. Some are devoted to the determination of the mobility of
vortices moving in a strip geometry in order to investigate the
influence of this motion on the generation of
noise~\cite{theunissen96,stan04}, others explore the vortex flow in
modulated channels designed by pinned vortices
\cite{anders00,besseling99,besseling05,kokubo02,kokubo04}. In such
systems, the vortex mobility depends upon the adequacy between the
inter-vortices distance and the periodicity of the pinned vortices
array and/or the channel width. For instance, it is well known that
this mobility drops when the vortices file diffuses into a channel
with a commensurate configuration of the pinning sites and that it
is very sensitive to the accuracy of this matching. Experimentally,
these effects have essentially been observed at a macroscopic scale
at which these matching effects result in an important increase of
the critical current for some precise values of the magnetic field.
The high sensitivity of the critical current towards the matching of
the characteristic lengths suggests that fluctuations in the channel
modulation, originating in the small movements of trapped vortices
for instance, could subtly influence the resulting flow from the
mobile vortices. From the theoretical point of view, the Langevin
equations describing the movements of particles located in a file
and diffusing in a fluctuating modulated channel are non linear and
their analytical solution is out of reach. Numerical studies have
also been performed~\cite{besseling99,kokubo02,kokubo04,besseling05}
; however, a complete description of such a behavior introducing
these fluctuations is still lacking.

In this letter we evidence that these fluctuations induce an
important enhancement of the vortices diffusion, one order of
magnitude higher than in the case of a bald channel. To exhibit this
very surprising result, we have used an original experimental
approach which allows to follow the movement of each particle
located in the file and determine its diffusion. More precisely, we
have compared the diffusion of particles moving in a fluctuating
modulated channel to the one observed when the same
particles are distributed in a pipe without modulation, hereafter called bald pipe.\\

In these experiments, millimetric stainless-steel balls are located
on the bottom electrode of an horizontal plane condenser, while a
metallic frame intercalated between the two electrodes and in
contact with the bottom electrode confines them. Depending on the
experiments, this frame is a disk whose external diameter is 10 mm
or a circular bald pipe, its external diameter of 10 mm and its
width of 2 mm preventing any crossing between particles
(Fig.\ref{fig:fig1}). When a voltage $V = 1$ kV is applied between
the two electrodes, the balls become monodispersely charged, repel
by each other and spreading throughout the whole available space. We
have shown that their electrostatic interaction is described by a
modified Bessel function of the second kind $K_0$ with a screening
length $\lambda=$ 0.48 mm. Notice that this interaction is exactly
similar to the inter-vortex interaction in superconductors
~\cite{galatola06,stjean01}. To introduce thermal noise, the whole
cell is fixed on loudspeakers supplied by a white noise voltage,
this mechanical shaking simulating a tunable effective temperature
that is measured in-situ~\cite{coupier05}. Throughout the
experiments, images of the particles are recorded in real time using
a camera. The interval between two successive snapshots is 150 ms
and five series of 10000 images have been recorded for each
experiment. With this choice, relevant statistics for the long-time
behavior of the displacements are obtained, the length of one
experiment remaining reasonable since the effective thermal bath is
characterized by a relaxation time of about 100 ms~\cite{coupier06}.
The diffusion of the particles is measured through the evolution
with time of their mean square displacements (m.s.d.) $\Delta
\theta^2(t)$ given by :
\begin{equation}
\Delta \theta^2(t)=\langle \big[\theta(t+t_0)-\theta(t_0)-\langle
\theta(t+t_0)-\theta(t_0)\rangle\big]^2\rangle,
\end{equation}
where the orthoradial coordinate $\theta$ (in radians) is the
cumulated angle, and not the modulo $2\pi$ angle in order to explore
an unbounded motion. The brackets $\langle\,\rangle$ denote ensemble
averaging over the initial time $t_0$ and a set of
statistically independent trajectories.\\

\begin{figure}
\resizebox{\columnwidth}{!}{\includegraphics{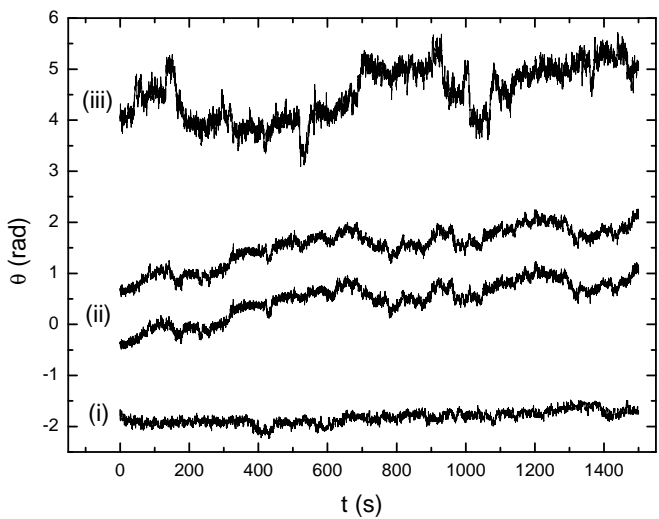}} \caption{
\label{fig:fig3}Evolution with time of the orthoradial coordinate
$\theta$ of a ball in : a bald circular pipe (i), a Wigner island
$N=19$ outer shell (two different balls) (ii), a Wigner island
$N=19$ inner shell (iii).}
\end{figure}

\begin{figure}
\resizebox{\columnwidth}{!}{\includegraphics{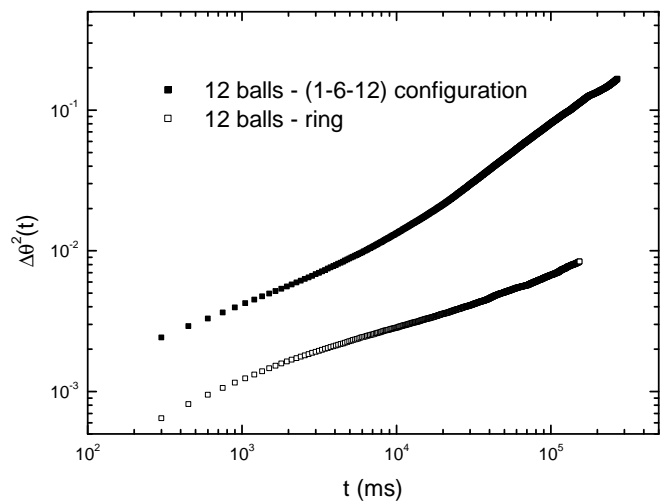}}
\caption{\label{fig:fig2}Comparison between the mean square
orthoradial displacements $\Delta \theta^2(t)$ of a ball moving in
the Wigner island $N=19$ outer shell and in a bald circular pipe
(log-log scale).}
\end{figure}

\begin{figure*}[t!]
\resizebox{2\columnwidth}{!}{\includegraphics{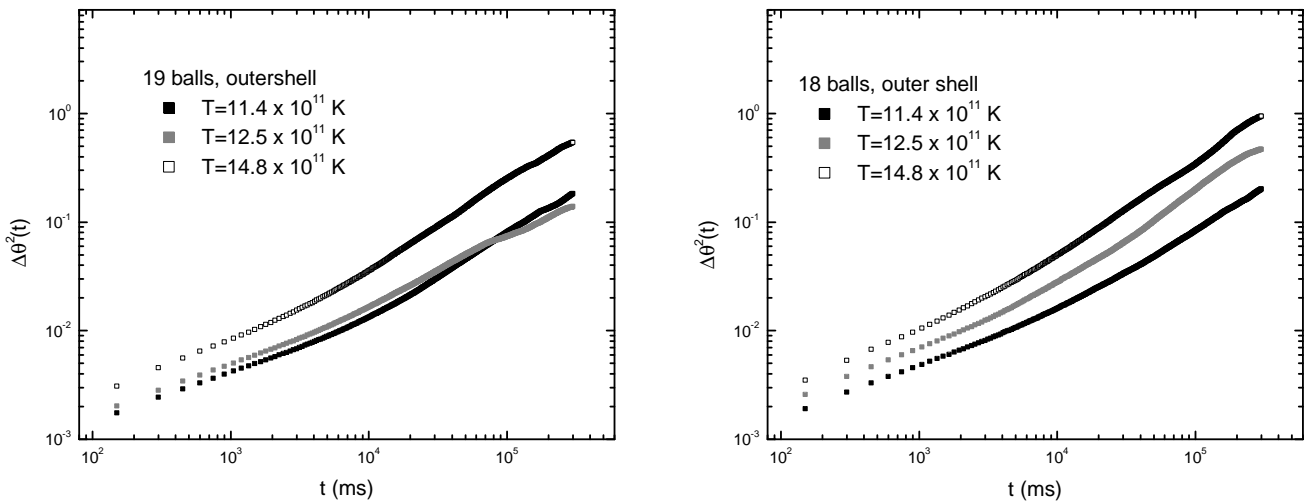}}
\caption{Evolution with time of the mean square orthoradial
displacement of a ball located in the outer shell of the
commensurate Wigner island $N=19$ and of the incommensurate Wigner
island $N=18$ (log-log scale).} \label{fig:fig4}
\end{figure*}

In order to observe a 1D movement in a fluctuating modulated
potential, we first considered the outer circular shell of a Wigner
island, constituting of 19 interacting particles confined in a
circular disk, for which the ground configuration corresponds to
self-organized pattern constituted of three concentric shells filled
by 1, 6 and 12 balls. This configuration is denoted (1-6-12)
hereafter (Fig.\ref{fig:fig1})~\cite{stjean01}. To observe well-
defined shells, the experiments have been performed at an effective
temperature equal to roughly a few tens of the inter-ball
interaction ; for this temperature range, the balls seldom jump from
one shell to another~\cite{coupier05}. The effective thermal
agitation only induces orthoradial motion while the radial
displacements are reduced as it can be seen on the balls
trajectories presented in the insets of
Fig.\ref{fig:fig1}~\cite{coupier05}. This offers the opportunity to
perform long-time orthoradial diffusion experiments. In the outer
shell, the relative mean orthoradial displacements of two
neighboring balls are about 0.07 radian (resp. 0.15 in the inner
shell), much smaller than the angular distance $2\pi/12$ (resp.
$2\pi/6$) between them. Thus, each shell can be considered as a
periodic ring. This ring presents global angular movements which are
coherently followed by the balls as shown in Fig.~\ref{fig:fig3}
(ii) where the trajectories of two different balls of the outer
shell are reported. Their global orthoradial movements are similar
but their trajectories present differences of low relative amplitude
at the small time-scale. Therefore, the outer twelve-ball shell is a
well-adapted realization of a periodic system of particles moving in
a fluctuating modulated potential due to the inner shell. Along the
orthoradial direction, the angular displacements of the balls have a
gaussian distribution. The evolution with time of the corresponding
m.s.d. $\Delta\theta^2(t)$ is presented in Fig.~\ref{fig:fig2} for
150 ms $\le t \le$ 150 s.  This angular m.s.d. increases with time
with a $t^\alpha$ dependance where $\alpha$ is close to 0.6.

We have compared these results with those obtained for the same
twelve balls moving in a circular bald pipe, the effective
temperature and the inter-ball interaction remaining the same
(Fig.\ref{fig:fig1}). We emphasize that, even if the movement of the
balls is not strictly 1D, no coupling effects resulting from the
circular channel geometry were observed. For instance, the
orthoradial movement of a single ball in this circular gutter is a
free diffusion~\cite{coupier06}. As in the Wigner island case, the
distribution of the angular movement of the balls is gaussian.
Long-time behavior of the angular m.s.d. differs dramatically from
the Wigner island case. Instead of $t^{\alpha}$ with $\alpha= 0.6$,
the movement is subdiffusive and characterized by a growth much
slower than in the modulated case, $\alpha$ varying from $1/2$ until
$\alpha= 0.4$ at the end of the experiment (the decrease of $\alpha$
being even more important at higher temperature). Equally remarkable
is that the amplitude of the diffusion is much smaller in this bald
pipe case than in the case of a fluctuating modulated channel ; 10
times smaller for instance for the last decade. This important
difference in the mobilities can be also directly observed on the
various trajectories presented on Fig.\ref{fig:fig3} (i-ii). The
role of the periodicity is confirmed when considering the diffusion
in the inner shell, which is seven times higher than in the outer
shell, with the same power law. Whereas an outer ball moves in a
modulated potential with a $2\pi/6$ period, an inner ball is indeed
submitted to a potential with a period twice as small  ; thus the
impulsion transfer that enhances the diffusion is linked to the
modulation of the potential.

These behaviors must be discussed in the frame of the Single File
Diffusion theory~\cite{harris65,vanbeijeren83} which describes the
diffusion of particles in a single channel where the crossings are
forbidden. For infinite systems with contact interaction, this
theory predicts that correlations between particles induce a
subdiffusion with $\alpha = 0.5$ whereas classical free diffusion is
characterized by $\alpha = 1$. We suggested that the smaller value
of $\alpha$ for the diffusion in the bald pipe resulted from the
cyclicity of the system associated to the non linearity of the
interaction~\cite{coupier06}. From this point of view, the $t^{0.6}$
behavior observed in the outer shell of the Wigner island is all the
more striking. Understanding this phenomenon and explaining why the
amplitude in a fluctuating modulated channel is higher than in a
smooth one is a complex issue, as it is well-known that the
diffusion of particles in a static modulated potential is slower
than for a free diffusion. Here the Single File Diffusion and the
modulation effects are coupled, and there is no theory describing
this coupling yet. Nevertheless, some propositions can be given. In
a recent paper, Bandyopadhyay \textit{et al.}~\cite{bandyopadhyay06}
have shown that diffusion of a single particle in a modulated
potential can be largely enhanced if the particle is excited by a
rapid fluctuating force. In the same way, we can attribute the
diffusion increase to the fluctuations of the modulated potential
felt by the outer balls, the fluctuating part of the force being
associated with the impulsion transfer resulting from the
oscillations of the balls
located in the neighbor inner shell.\\

To reveal how the channel modulation and their fluctuations can play
a role, we have performed diffusion experiments in which we have
modified the inter-shell interaction without drastically changing
the moving file. We have compared the diffusion of the outer shell
of the $N=19$ system to the one obtained with a $N = 18$ island
whose ground configuration is (1-6-11). The advantage of these two
systems is that they roughly have the same number of balls in the
outer shell whereas their relative symmetries strongly differ : the
commensurate system $N=19$ exhibits a three fold symmetry whereas
the $N = 18$ system is incommensurate. This difference of relative
symmetry between the two shells modify the inter-shell coupling and
thus might influence the effect of this coupling on the diffusion.
The variations of the corresponding $\Delta\theta^2(t)$ are
presented in Fig.\ref{fig:fig4} for three different temperatures
$T$. For long times, the m.s.d. associated with the incommensurate
$N = 18$ system increases roughly with the same power law ($\alpha =
0.6$) as for the commensurate island $N=19$. However, the evolution
with $T$ of the diffusion amplitudes differ according to $N$ and a
detailed analysis of these differences proves that it is greatly
influenced by the relative inter-shell movement. Indeed, we showed
in a previous paper that for $11\times10^{11}$ K$\le T
\le12\times10^{11}$ K, the relative shells' movements are small in
the $N = 19$ system~\cite{coupier05}. In the diffusion experiments,
one can observe that, in the same range of temperature, the
diffusion curves are remarkably roughly superposable
(Fig.\ref{fig:fig4}). Likewise, at higher temperature, we observe a
stronger diffusion as the inter-shell movement is increased.
Furthermore, for the same temperatures, the shells for $N = 18$ are
always unlocked~\cite{coupier05} and the amplitudes of the
inter-shell displacements grow with temperature. Similar behaviors
can be observed on the diffusion curves : whatever the temperature,
the corresponding diffusion curves are always well separated and the
diffusion amplitude increases also with the effective temperature
(Fig.\ref{fig:fig4}). Then we can conclude that there exists a
strong correlation between the inter-shell movement and the
diffusion in each shell.

However, it would be naive to picture the two shells as a set of
gear-wheels, the correlation effect being undoubtedly more subtle.
Indeed, a gear-wheels picture could suggest that a more important
diffusion increase would be obtained  in the case of two
commensurate shells but this assumption does not correspond to the
observations. Indeed, whatever the effective temperature, diffusion
for balls of $N = 18$ is always higher than for $N = 19$, the
maximum difference being  precisely at the temperature at which the
$N=19$ system is locked whereas the $N=18$ system is unlocked
(Fig.\ref{fig:fig4}). When the temperature increases, the
differences are less important, the file of particles can be
considered as floating above the modulated potential. It then only
feels an averaged effect where the influence of the peculiar
symmetries is hidden. We stress that a similar effect of the
commensurability can also be observed for balls in the inner shells.
Thus, even though the fluctuations of the modulated potential
induces impulsion transfers, a better transfer is obtained in
incommensurate configurations, a too good matching of the periods
inducing the locking of the whole system rather to a positive cooperation.\\

In conclusion, the diffusion of a file of particles can be
accelerated when it moves in a fluctuating modulated potential. This
enhancement presents two strong characteristics : the orthoradial
m.s.d. increases as a power law $t^\alpha$ with $\alpha>$ 0.5
whereas the diffusion in a circular Single File Diffusion is
characterized by $\alpha\le 0.5$. Moreover, the diffusion amplitude
in a fluctuating potential is much larger than in a bald pipe. This
effect seems to be favored by the incommensurability between the
modulation of the potential and the periodicity of the moving file.
This effect could be very important in the case of mobile vortices
moving between pinned ones in superconducting devices. In
particular, it has to be taken into account to design well adapted
patterns of traps in order to optimize the critical current. From
the theoretical point of view, getting a better understanding of
this phenomenon would require to develop a complete SFD theory
including a fluctuating potential.

\end{document}